# Computing Elementary Symmetric Polynomials with a Sublinear Number of Multiplications

*Preliminary version*

Vince Grolmusz *

**Abstract**

Elementary symmetric polynomials $S_n^k$ are used as a benchmark for the bounded-depth arithmetic circuit model of computation. In this work we prove that $S_n^k$ modulo composite numbers $m = p_1 p_2$ can be computed with much fewer multiplications than over any field, if the coefficients of monomials $x_{i_1} x_{i_2} \cdots x_{i_k}$ are allowed to be 1 either mod $p_1$ or mod $p_2$ but not necessarily both. More exactly, we prove that for any constant $k$ such a representation of $S_n^k$ can be computed modulo $p_1 p_2$ using only $\exp(O(\sqrt{\log n} \log \log n))$ multiplications on the most restricted depth-3 arithmetic circuits, for $\min(p_1, p_2) > k!$. Moreover, the number of multiplications remain sublinear while $k = O(\log \log n)$. In contrast, the well-known Graham-Pollack bound yields an $n - 1$ lower bound for the number of multiplications even for the exact computation (not the representation) of $S_n^2$. Our results generalize for other non-prime power composite moduli as well. The proof uses the famous BBR-polynomial of Barrington, Beigel and Rudich.

## 1 Introduction

Surprising ideas sometimes lead to considerable improvements in algorithms even for the simplest computational tasks, let us mention here the integer-multiplication algorithm of Karatsuba and Ofman [KO63] and the matrix-multiplication algorithm of Strassen [Str69].

A new field with surprising algorithms is quantum computing. The most famous and celebrated results are Shor's algorithm for integer factorization [Sho97] and Grover's database-search algorithm [Gro96].

Since realizable quantum computers can handle only very few bits today, there are no practical applications of these fascinating quantum algorithms.

Computations involving composite, non-prime-power moduli (say, 6), on the other hand, can actually be performed on any desktop PC, but, unfortunately, we have only little evidence on the power or applicability of computations modulo composite numbers (see, e.g., the circuit given by Kahn and Meshulam [KM91], or the low-degree polynomial of Barrington, Beigel and Rudich [BBR94]).

One of the problems here is the interpretation of the output of the computation. Several functions are known to be hard if computed modulo a prime. If we compute the same function $f$ with 0-1 values modulo 6, then it will also be computed modulo - say - 3, since $f(x) \equiv 1 \pmod 6 \Longrightarrow f(x) \equiv 1 \pmod 3$ and $f(x) \equiv 0 \pmod 6 \Longrightarrow f(x) \equiv 0 \pmod 3$, consequently, computing $f$ this way cannot be easier mod 6 than mod 3. This difficulty is circumvented in a certain sense by the definition of the weak representation of Boolean functions by mod 6 polynomials, defined in [TB98] and [BBR94].

*Department of Computer Science, Eötvös University, Budapest, Pázmány P. stny. 1/C, H-1117 Budapest, Hungary; E-mail: grolmusz@cs.elte.hu

We will consider here another interpretation of the output, called a-strong representation (Definition 1). This definition will be more suitable for computations, where the output is a polynomial and not just a number.

Our goal is to compute elementary symmetric polynomials

$$S_n^k = \sum_{\substack{I \subset \{1,2,\ldots,n\} \\ |I|=k}} \prod_{i \in I} x_i \tag{1}$$

modulo non-prime-power composite numbers with a much smaller number of multiplications than it is possible over rationals or prime moduli.

Our model of computation is the arithmetic circuit model of depth 3, circuits in this model are often called $\Sigma\Pi\Sigma$ circuits [RSV00], [Shp].

$\Sigma\Pi\Sigma$ circuits perform computations of the following form:

$$\sum_{i=1}^{r} \prod_{j=1}^{s_i} (a_{ij1}x_1 + a_{ij2}x_2 + \cdots + a_{ijn}x_n + b_{ij}).$$

If all the $b_{ij} = 0$ and all the $s_i$'s are the same number, then the circuit is called a homogeneous circuit, otherwise it is inhomogeneous. The size of the circuit is the number of gates in it: $1 + r + \sum_{i=1}^{r} s_i$.

A special class of homogeneous $\Sigma\Pi\Sigma$ circuits is called in [RSV00] *the graph model*: here all $s_i = 2$ and all $a_{ij\ell}$ coefficients are equal to 1, and, moreover, the clauses of a product cannot contain the same variable twice. Consequently, such a product corresponds to a complete bipartite graph on the variables as vertices.

Graham and Pollack [GP72] asked that how many edge-disjoint bipartite graphs can cover the edges of an $n$-vertex complete graph. They proved that $n-1$ bipartite graphs are sufficient and necessary. Later, Tverberg gave a very nice proof for this statement [Tve82]. Having relaxed the disjointness-property, Babai and Frankl [BF92] asked that what is the minimum number of bipartite-graphs, which covers every edge of an $n$-vertex complete graph by an odd multiplicity. Babai and Frankl proved that $(n-1)/2$ bipartite graphs are necessary. The optimum upper bound for the odd-cover was proved by Radhakrishnan, Sen and Vishwanathan [RSV00]. Radhakrishnan, Sen and Vishwanathan also gave matching upper bounds for covers, when the off-diagonal elements of matrix $M$ are covered by multiplicity 1 modulo a prime.

By a result of Ben-Or [Shp], every elementary symmetric polynomial $S_n^k$ (and similarly, every symmetric function) can be computed over fields by size-$O(n^2)$ inhomogeneous $\Sigma\Pi\Sigma$ circuits, using one-variable polynomial interpolation. This result shows the power of arithmetic circuits over Boolean circuits with MOD $p$ gates, since as it was proved by Razborov [Raz87] and Smolensky [Smo87] that MAJORITY – a symmetric function – needs exponential size to be computed on any bounded-depth Boolean circuits.

Note, that our construction with homogeneous circuits beats this bound for small $k$'s.

$S_n^k$ can be naturally computed by $\binom{n}{k}$ product-gates by a homogeneous $\Sigma\Pi\Sigma$ circuit over any ring by the circuit of (1).

Nisan and Wigderson [NW97] showed that any homogeneous $\Sigma\Pi\Sigma$ circuit needs size $\Omega((n/2k)^{k/2})$ for computing $S_n^k$. This result shows that the homogeneous circuits are much weaker in computing elementary symmetric polynomials than the inhomogeneous ones. Nisan and Wigderson also examined bilinear and multi-linear circuits in [NW97]. Note that the circuits in our constructions for $S_n^2(x,y)$ and for $S_n^k(x^1, x^2, \ldots, x^k)$ are also multi-linear circuits.

We should note, that exponential lower bounds were proved recently for simple functions for $\Sigma\Pi\Sigma$ circuits by Grigoriev and Razborov [GR00] and by Grigoriev and Karpinski [GK98].

Most recently, Raz and Shpilka got nice lower bound results for constant-depth arithmetic circuits [RS01], and Raz [Raz02] proved a $\Omega(n^2 \log n)$ lower bound for matrix-multiplication in this model, solving a long-standing open problem.

## 1.1 Alternative strong representation of polynomials

Several authors (e.g., [TB98], [BBR94]) defined the weak and strong representations of Boolean functions for integer moduli. Here we need the definition of a sort of strong representation of polynomials modulo composite numbers. We call this representation *alternative-strong* representation, abbreviated *a-strong representation*:

**Definition 1** *Let $m$ be a composite number $m = p_1^{e_1} p_2^{e_2} \cdots p_\ell^{e_\ell}$. Let $Z_m$ denote the ring of modulo $m$ integers. Let $f$ be a polynomial of $n$ variables over $Z_m$:*

$$f(x_1, x_2, \ldots, x_n) = \sum_{I \subset \{1,2,\ldots,n\}} a_I x_I,$$

*where $a_I \in Z_m$, $x_I = \prod_{i \in I} x_i$. Then we say that*

$$g(x_1, x_2, \ldots, x_n) = \sum_{I \subset \{1,2,\ldots,n\}} b_I x_I,$$

*is an* a-strong representation *of $f$ modulo $m$, if*

$$\forall I \subset \{1, 2, \ldots, n\} \ \exists j \in \{1, 2, \ldots, \ell\}: \quad a_I \equiv b_I \pmod{p_j^{e_j}},$$

*and if for some $i$, $a_I \not\equiv b_I \pmod{p_i^{e_i}}$, then $b_I \equiv 0 \pmod{p_i^{e_i}}$.*

**Example 2** *Let $m = 6$, and let $f(x_1, x_2, x_3) = x_1 x_2 + x_2 x_3 + x_1 x_3$, then $g(x_1, x_2,, x_3) = 3 x_1 x_2 + 4 x_2 x_3 + x_1 x_3$ is an a-strong representation of $f$ modulo 6.*

Note, that the earlier (strong-, weak-) representations of functions contained constraints for the *value* of certain functions. Now we are requiring that the *form* of the representation satisfy modular constraints.

Our goal in this work is to show that the elementary symmetric polynomials have a-strong representations modulo composites which can be computed by much smaller homogeneous $\Sigma\Pi\Sigma$ arithmetic circuits than the original polynomial.

Unfortunately, we cannot hope for such results for all multivariate polynomials, as it is shown by the next Theorem:

**Theorem 3** *Let*

$$f(x_1, x_2, \ldots, x_n, y_1, y_2, \ldots, y_n) = \sum_{i=1}^n x_i y_i$$

*the inner product function. Suppose that a $\Sigma\Pi\Sigma$ circuit computes an a-strong representation of $f$ modulo 6. Then the circuit must have at least $\Omega(n)$ multiplication gates.*

**Proof:** Let $g$ be the a-strong representation of $f$. Then in $g$, at least the half of monomials $x_i y_i$ has coefficients equal to 1 modulo either 2 or 3. Without restricting the generality, let us assume that monomials $x_1 y_1, x_2 y_2, \ldots, x_{\lceil n/2 \rceil} y_{\lceil n/2 \rceil}$ have coefficients 1 modulo 3. When we compute $g$ modulo 6 we will learn also the inner product of two vectors modulo 3, each consisting of the first $\lceil n/2 \rceil$ variables. It is well known that the communication complexity of computing the inner product mod 3 is $\Omega(n)$ (see e.g., [Gro95]).

Since arithmetic $\Sigma\Pi\Sigma$ circuits modulo 6 with $u$ multiplication-gates of in-degree 2 can be evaluated by a 2-party communication protocol using only $O(u)$ bits, we get: $u = \Omega(n)$. □

## 2 Our Constructions

First we construct a-strong representations with a small number of multiplications for the following polynomial:

$$S_n^2(x,y) = \sum_{\substack{i,j \in \{1,2,\ldots,n\} \\ i \neq j}} x_i y_j, \tag{2}$$

and for $x = y$ we will get that $2S_n^2(x) = S_n^2(x,x)$, and this will imply our result for any composite, odd, non-prime-power moduli $m$:

**Theorem 4** *(i) Let $m = p_1 p_2$, where $p_1 \neq p_2$ are primes. Then an a-strong representation of $S_n^2(x,y)$ modulo $m$ can be computed on a homogeneous $\Sigma\Pi\Sigma$ circuit of size*

$$\exp(O(\sqrt{\log n \log \log n})).$$

*(ii) Let the prime decomposition of $m = p_1^{e_1} p_2^{e_2} \cdots p_r^{e_r}$. Then an a-strong representation of $S_n^2(x,y)$ modulo $m$ can be computed on a homogeneous $\Sigma\Pi\Sigma$ circuit of size*

$$\exp\left(O\left(\sqrt[r]{\log n (\log \log n)^{r-1}}\right)\right).$$

**Corollary 5** *(i) Let $m = p_1 p_2$, where $p_1 \neq p_2$ are odd primes. Then an a-strong representation of the second elementary symmetric polynomial $S_n^2(x)$ modulo $m$ can be computed on a homogeneous $\Sigma\Pi\Sigma$ circuit of size*

$$\exp(O(\sqrt{\log n \log \log n})).$$

*(ii) Let the prime decomposition of the odd $m$ be $m = p_1^{e_1} p_2^{e_2} \cdots p_r^{e_r}$. Then an a-strong representation of the second elementary symmetric polynomial $S_n^2(x)$ modulo $m$ can be computed on a homogeneous $\Sigma\Pi\Sigma$ circuit of size*

$$\exp\left(O\left(\sqrt[r]{\log n (\log \log n)^{r-1}}\right)\right).$$

Since the $\Sigma\Pi\Sigma$ circuit in our construction correspond to the graph-model [RSV00], we have the following graph-theoretical corollary, showing a cover with much fewer bipartite graphs than in the linear lower bound of Graham and Pollack:

**Corollary 6** *For any $m = p_1^{e_1} p_2^{e_2} \cdots p_r^{e_r}$, there exists an explicitly constructible bipartite cover of the edges of the complete n-vertex-graph, such that for all edges e there exists an $i : 1 \leq i \leq r$, that the number of the bipartite graphs, covering e is congruent to 1 modulo $p_i^{e_i}$. Moreover, the total number of the bipartite graphs in the cover is*

$$\exp\left(O\left(\sqrt[r]{\log n (\log \log n)^{r-1}}\right)\right).$$

The following theorem gives our result for general $k$. Our goal is to compute an a-strong representation of polynomials $S_n^k(x)$ for $n \geq k \geq 2$. Let us first define

$$S_n^k(x^1, x^2, \ldots, x^k) = \sum_{i_1, i_2, \ldots, i_k} x_{i_1}^1 x_{i_2}^2 \cdots x_{i_k}^k,$$

where the summation is done for all $k!$ orders of all $k$ element subsets $I = \{i_1, i_2, \ldots, i_k\}$ of $\{1, 2, \ldots, n\}$, and $x^i = (x_1^i, x_2^i, \ldots, x_n^i)$, for $i = 1, 2, \ldots, k$.

**Theorem 7** *Let $m = p_1^{e_1} p_2^{e_2} \cdots p_r^{e_r}$. Then an a-strong representation of $S_n^k(x^1, x^2, \ldots, x^k)$ modulo m can be computed on a homogeneous and multi-linear $\Sigma\Pi\Sigma$ circuit of size*

$$\exp\left(\exp(O(k))\sqrt[r]{\log n}(\log \log n)\right).$$

Note, that this circuit-size is sublinear in $n$ for any constant $k$ and for large enough $n$. Moreover, the sublinearity holds while $k < c \log \log n$, for a small enough $c > 0$.

For moduli $m$, relative prime to $k!$, this implies:

**Corollary 8** *If m is relative prime to $k!$, then an a-strong representation of $S_n^k(x)$ modulo m can be computed on a homogeneous $\Sigma\Pi\Sigma$ circuit of size*

$$\exp\left(\exp(O(k))\sqrt[r]{\log n}(\log \log n)\right).$$

## 2.1 The construction for computing $S_n^2$

**Proof:** Note, that $S_n^2(x, y)$ contains the sum of the monomials $x_i y_j$ for all $i \neq j$. Let us arrange these monomials as follows: Let $x_i'$s and $y_j'$s be assigned to the rows and columns of an $n \times n$ matrix $M$, and the position in row $i$ and column $j$ contains monomial $x_i y_j$:

$$M = \begin{pmatrix} & y_1 & y_2 & \cdots & y_n \\ x_1 & x_1 y_1 & x_1 y_2 & \cdots & x_1 y_n \\ x_2 & x_2 y_1 & x_2 y_2 & \cdots & x_2 y_n \\ \vdots & \vdots & \vdots & \ddots & \vdots \\ x_n & x_n y_1 & x_n y_2 & \cdots & x_n y_n \end{pmatrix} \tag{3}$$

Any product of the form

$$(x_{i_1} + x_{i_2} + \cdots + x_{i_v})(y_{j_1} + y_{j_2} + \cdots + y_{j_w}) \tag{4}$$

naturally corresponds to a $v \times w$ submatrix of matrix $M$. We call these submatrices rectangles. Clearly, any a-strong representation modulo $m$ of polynomial $S_n^2(x, y)$ can be got from a cover of matrix $M$ by rectangles of the form (4), satisfying the following properties:

**Property (a):** The number of the rectangles covering any elements of the diagonal is a multiple of $m$;

**Property (b):** Any non-diagonal element $x_i y_j$ of $M$ is covered by $d_{ij}$ rectangles, where either $d_{ij} \equiv 1 \pmod{p_1}$ or $d_{ij} \equiv 0 \pmod{p_1}$ and $d_{ij} \equiv 1 \pmod{p_2}$.

Clearly, a (bilinear) $\Sigma\Pi\Sigma$ circuit compute an a-strong representation of polynomial $S_n^2(x,y)$ if and only if when the corresponding rectangle-cover satisfies Properties (a) and (b). The construction of such a low-cardinality rectangle cover is implicit in papers [Gro00a] and [Gro00b]. We present here a short direct proof which is easily generalizable for proving the results in the next section for higher dimensional matrices.

Rectangles, covering $M$, will be denoted

$$R(I,J) = \left(\sum_{i \in I} x_i\right)\left(\sum_{j \in J} y_j\right).$$

We define now an initial cover of the non-diagonal elements of $M$ by rectangles.

Let $N = \lceil \log n \rceil$, and for $1 \leq i, j \leq n$, let $i = (i_1, i_2, \ldots, i_g)$ and $j = (j_1, j_2, \ldots, j_g)$ denote their $N$-ary forms (i.e., $0 \leq i_t, j_t \leq N-1$, for $t = 1, 2, \ldots, g$, where $g = \lceil \log_N(n+1) \rceil$.)

Then let us define for $t = 1, 2, \ldots, g$ and $\ell = 0, 1, \ldots, N-1$:

$$I_t^\ell = \{i : i_t = \ell\}, \quad J_t^\ell = \{j : j_t \neq \ell\}.$$

Now consider the cover given by the following rectangles:

$$R(I_t^\ell, J_t^\ell) : t = 1, 2, \ldots, g, \ \ell = 0, 1, \ldots, N-1.$$

Now, in this cover, any element $x_i y_j$ of $M$ will be covered by $H_N(i,j)$-times, where $H_N(i,j)$ stands for the Hamming-distance of the N-ary forms of $i$ and $j$, that is, at most $g$-times. Note, that the diagonal elements are not covered at all, so Property (a) is satisfied, while Poperty (b) typically not.

The total number of covering rectangles is $h = gN = O((N \log n)/\log N)$.

Now, our goal is to turn this cover to another one, which already satisfies not only Property (a), but also Property (b). For this transformation we need to apply a multivariate polynomial $f$ to our rectangle-cover in a very similar way as we applied polynomials to set-systems in [Gro01] and to codes in [Gro02].

**Definition 9** *Let $R_1, R_2, \ldots, R_h$ be a rectangle-cover of a matrix $M = \{x_i y_j\}$, and let $f$ be a $h$-variable multi-linear polynomial written in the following form:*

$$f(z_1, z_2, \ldots, z_h) = \sum_{K \subset \{1,2,\ldots,h\}} a_K z_K,$$

*where $0 \leq a_K \leq m-1$ are integers, and $z_K = \prod_{k \in K} z_k$. Then the $f$-transformation of the rectangle-cover $R_1 R_2, \ldots, R_h$ contains $\sum_{K \subset \{1,2,\ldots,h\}} a_K$ rectangles, each corresponding to a monomial of $f$. $z_K = \prod_{k \in K} z_k$ is corresponded to the (possibly empty) rectangle of $\bigcap_{k \in K} R_k$.*

Note, that another way of interpreting this definition is as follows: the variables $z_k$ correspond to the rectangles of the cover, and if we imagine the rectangles filled with 1's, then the product of the variables, i.e., the monomials, correspond to the Hadamard-product of the corresponding all-1 rectangles, resulting an all-1 rectangle, which, in turn, equals to their intersection.

Note also, that polynomial $f$ is, in fact, considered over the ring $Z_m$, with a fixed (small) representation of its coefficients from the set of integers.

**Lemma 10** *Let $u^{ij} \in \{0,1\}^h$ characterize the rectangle-cover of the entry $x_i y_j$ of matrix $M$ as follows:*

$$R_s \text{ covers } x_i y_j \iff u^{ij}_s = 1.$$

*Then entry $x_i y_j$ is covered by exactly $f(u^{ij})$ rectangles from the $f$-transformation of the rectangle-cover $R_1, R_2, \ldots, R_h$*

**Proof:** In $f(z)$, exactly those monomials $z_K$ contributes 1 to the value of $f(u^{ij})$ whose variables are all-1 in vector $u^{ij}$. This happens exactly when $u^{ij}_k = 1$ for all $k \in K$, that is, $x_i y_j$ is covered by the intersection of rectangles $\bigcap_{k \in K} R_k$. □

The proof of the following lemma is obvious:

**Lemma 11** *The intersection of finitely many rectangles is a (possibly empty) rectangle. Any rectangle, covering a part of matrix $M$ of (3) corresponds to a single (bilinear) multiplication.*

□

It remains to prove that there exists an $f$, with a small number of monomials, and with properties which leads to a cover, satisfying Properties (a) and (b). We will use the famous BBR polynomial of Barrington, Beigel and Rudich [BBR94]:

**Theorem 12 (Barrington, Beigel, Rudich)** *Let $m = p_1^{e_1} p_2^{e_2} \cdots p_r^{e_r}$. For any integers $d, \ell$, $1 \leq d \leq \ell$ there exists an $f_{d,\ell}$ explicitly constructible, $\ell$-variable, degree-$O(d^{1/r})$ multilinear polynomial with coefficients from $Z_m$, such that*

(i) *for any $z \in \{0,1\}^\ell$, which contains at most $d$ 1's:*

$$f_{d,\ell}(z) \equiv 0 \pmod{m} \iff z = 0,$$

(ii) *If $f_{d,\ell}(z) \not\equiv 0 \pmod{m}$, then there exists $i \in \{1, 2, \ldots, r\}$: $f_{d,\ell}(z) \equiv 1 \pmod{p_i^{e_i}}$, and if $f_{d,\ell}(z) \not\equiv 1 \pmod{p_j^{e_j}}$, then $f_{d,\ell}(z) \equiv 0 \pmod{p_j^{e_j}}$.*

**Proof:** The proof of (i) is given in [BBR94]. The proof of part (ii) is obvious for $e_i = 1$ from the little Fermat-theorem, and from a lemma of Beigel and Tarui [BT94] (for modulus-amplifying), in general. □

Now we can prove Theorem 4 part (i), the proof of part (ii) remains to the full version. Let $m = p_1 p_2$, let $\ell = h = gN$, $d = g$. Then $f_{g,gN}$ has

$$\binom{h}{O(\sqrt{g})} \tag{5}$$

monomials. Consequently, if we transform our cardinality-$h$ rectangle cover by Definition 9 with polynomial $f_{g,gN}$, then the resulting cover satisfies Properties (a) and (b) and has cardinality (5). This implies an $\exp(O(\sqrt{\log n \log \log n}))$ cover. By Lemma 11, a $\Sigma\Pi\Sigma$ circuit is immediate with $\exp(O(\sqrt{\log n \log \log n}))$ multiplication-gates.
□

## 2.2 The construction in general

We describe a construction similarly as in the case $k = 2$. In this preliminary version we prove only the $m = p_1 p_2$, $r = 2$ case, the proof for general $m$ is analogous.

First, let $M' = \{m_{i_1,i_2,\ldots,i_k}\}$ be a $k$-dimensional analogon of $M$ of equation (3), that is, an $\overbrace{n \times n \times n \times \cdots \times n}^{k}$ matrix, where $m_{i_1,i_2,\ldots,i_k} = x^1_{i_1} x^2_{i_2} \cdots x^k_{i_k}$. Now we should again construct a cover of $M'$, this time with $k$-dimensional boxes, corresponding to the $k$-linear products

$$\prod_{i=1}^{k}(x^i_{j_{i1}} + x^i_{j_{i2}} + \cdots + x^i_{j_{i\ell_i}}),$$

satisfying that only those entries will be covered, which have no two equal indices, and the covering multiplicity of these entries should be non-zero modulo $m$.

First we need to define an initial box-cover of those entries of the $k$-dimensional matrix $M'$, which have no two identical indices.

For our proof it is very important, that this initial cover has low multiplicity: every covered element of $M'$ should be covered only by $O(\log n)$ $k$-dimensional boxes for constant $k$'s. The construction of such initial cover in the $k = 2$ case was quite easy, now we must use some more intricate approach.

Using a family of perfect hash functions (see e.g., [FK84]), for integers $n, k, b$: $2 \leq k \leq b = O(k)$, $k \leq n$, one can obtain a matrix $H(n, k, b) = \{h_{ij}\}$ with $u = \exp(O(k)) \log n$ rows and $n$ columns, with entries from the set $\{0, 1, \ldots, b-1\}$, such that for any $k$-element subset $J$ of the $n$ columns, there exists an $i : 1 \leq i \leq u$:

$$h_{ij} : j \in J$$

are pairwise different elements of the set $\{0, 1, \ldots, b-1\}$.

Matrix $H(n, k, b)$ will be used for the definition of our initial cover as follows:

For any $i : 1 \leq i \leq u$, and any $\sigma : \{1, 2, \ldots, k\} \to \{0, 1, \ldots, b-1\}$ injective function we define the $k$-dimensional box:

$$R(i, \sigma) = \{m_{j_1, j_2, \ldots, j_k} : h_{ij_1} = \sigma(1), h_{ij_2} = \sigma(2), \ldots, h_{ij_k} = \sigma(k)\}.$$

There are $u$ possible $i$'s and $k^{O(k)}$ possible $\sigma$'s, so there are $k^{O(k)} \log n$ boxes in this cover. Box $R(i, \sigma)$ covers only $m_{j_1, j_2, \ldots, j_k}$'s with pairwise different indices.

Any $m_{j_1, j_2, \ldots, j_k}$ with pairwise different indices is covered by exactly that many $k$-dimensional boxes, as the number of rows with pairwise different elements of the sub-matrix, containing column $j_1$, column $j_2$, ..., column $j_k$ of matrix $H(n, k, b)$. This number is at least 1 (from the perfect-hashing property) and at most $u$ (that is, the number of rows of $H(n, k, b)$).

Now, exactly as in the proof of the $S_n^2$ case, we apply the polynomial $f_{d,\ell}$ of Theorem 12 with $d = u$, $\ell = k^{O(k)} \log n$, to this box-cover by Definition 9 and by the higher-dimensional version of Lemma 10.

The result is a box-cover of cardinality $\exp(\exp(O(k))\sqrt{\log n \log \log n})$, proving Theorem 7, in case $m = p_1 p_2$.

**Acknowledgment.**

The author is grateful for Gábor Tardos for discussions on the subject. The author also acknowledges the partial support of János Bolyai Fellowship, and research grants EU FP5 IST FET No. IST-2001-32012, OTKA T030059 and an ETIK grant.